\documentclass[times,twocolumn,final]{elsarticle}

\usepackage{arxiv}
\usepackage{graphicx}%
\usepackage{multirow}%
\usepackage{amsmath,amssymb,amsfonts}%

\usepackage{amsthm}%
\usepackage{mathrsfs}%
\usepackage[title]{appendix}%
\usepackage{xcolor}%
\usepackage[utf8]{inputenc}
\usepackage{booktabs}%
\usepackage{algpseudocode,algorithm}
\MakeRobust{\Call}
\usepackage{pifont}
\usepackage{adjustbox}
\usepackage{graphicx}
\usepackage{float}
\usepackage{placeins}
\usepackage{fancyhdr}
\usepackage[hyphens]{url}
\usepackage{hyperref}
\usepackage{caption}

\usepackage{subcaption}
\usepackage{enumitem}
\usepackage[font=small,labelfont=bf,labelsep=period,skip=4pt]{caption} 
\usepackage{booktabs,tabularx,ragged2e,array,multirow,graphicx}
\DeclareUnicodeCharacter{2606}{\ding{73}}
\newcolumntype{Y}{>{\RaggedRight\arraybackslash}X}

\fancypagestyle{firstpagestyle}{
    \fancyhf{} 
    \fancyhead{} 
    \fancyfoot{} 
}

\fancypagestyle{default}{
    \fancyhf{}
    \fancyhead[R]{\thepage} 
}

\begin{document}


\title{Artificial Intelligence for the Assessment of Peritoneal Carcinosis
during Diagnostic Laparoscopy for Advanced Ovarian Cancer}

\fntext[equal1]{These authors contributed equally as first authors.}
\fntext[equal2]{These authors contributed equally as last authors.}

\author[1,2,3]{Riccardo \snm{Oliva}\fnref{equal1}}
\author[4]{Farahdiba \snm{Zarin}\fnref{equal1}}
\author[1,2]{Alice Zampolini \snm{Faustini}\fnref{equal1}}
\author[5]{Armine \snm{Vardazaryan}}
\author[1,2]{Andrea \snm{Rosati}}
\author[4,5]{Vinkle \snm{Srivastav}}
\author[6]{Nunzia Del \snm{Villano}}
\author[3]{Jacques \snm{Marescaux}}
\author[1,2]{Giovanni \snm{Scambia}}
\author[5,7]{Pietro \snm{Mascagni}\corref{corresp}}
\cortext[corresp]{Corresponding author: \texttt{pietro.mascagni@ihu-strasbourg.eu}}
\author[4,5]{Nicolas \snm{Padoy}\fnref{equal2}}
\author[1,2]{Anna \snm{Fagotti}\fnref{equal2}}

\address[1]{Gynecologic Oncology Unit, Fondazione Policlinico Universitario Agostino Gemelli IRCCS, Rome, Italy}
\address[2]{Università Cattolica del Sacro Cuore, Rome, Italy}
\address[3]{IRCAD, Research Institute against Digestive Cancer, Strasbourg, France}
\address[4]{University of Strasbourg, CNRS, INSERM, ICube, UMR7357, Strasbourg, France}
\address[5]{IHU Strasbourg, Strasbourg, France}
\address[6]{Università degli studi di Modena, Modena, Italy}
\address[7]{Bioimage Analysis Center, Fondazione Policlinico Universitario A. Gemelli IRCCS, Rome, Italy}

\received{XXX}
\finalform{XXX}
\accepted{XXX}
\availableonline{XXX}
\communicated{XXX}

\begin{abstract}

\textbf{Objective:} To develop an artificial intelligence (AI) system to assist intraoperative decision-making during diagnostic laparoscopy (DL) in patients with advanced ovarian cancer (AOC).

\textbf{Background:} AOC is often diagnosed at an advanced stage with peritoneal carcinosis (PC). Fagotti score (FS) assessment at DL guides treatment planning by estimating surgical resectability, but its subjective and operator-dependent nature limits reproducibility and widespread use.

\textbf{Methods:} Videos of patients undergoing DL with concomitant FS assessments at a referral center were retrospectively collected and divided into a development dataset, for data annotation, AI training and evaluation, and an independent test dataset, for internal validation. In the development dataset, FS-relevant frames were manually annotated for anatomical structures and PC. Deep learning models were trained to automatically identify FS-relevant frames, segment structures and PC, and predict video-level FS and indication to surgery (ItS). AI performance was evaluated using Dice score for segmentation, F1-scores for anatomical stations (AS) and ItS prediction, and root mean square error (RMSE) for final FS estimation.

\textbf{Results:} In the development dataset, the segmentation model trained on 7,311 frames, achieved Dice scores of 70±3\% for anatomical structures and 56±3\% for PC. Video-level AS classification achieved F1-scores of 74±3\% and 73±4\%, FS prediction showed normalized RMSE values of 1.39±0.18 and 1.15±0.08, and ItS reached F1-scores of 80±8\% and 80±2\% in the development (n=101) and independent test datasets (n=50), respectively. 

\textbf{Conclusions:} This is the first AI model to predict the feasibility of cytoreductive surgery providing automated FS estimation from DL videos. Its reproducible and reliable performance across datasets suggests that AI can support surgeons through standardized intraoperative tumor burden assessment and clinical decision-making in AOC.


\end{abstract}

\maketitle
\thispagestyle{firstpagestyle}

\section{Introduction}
\label{sec:introduction}

Advanced ovarian cancer (AOC) remains one of the most lethal gynecologic malignancies worldwide, largely because it is usually diagnosed at an advanced stage, often following extensive peritoneal dissemination and a high tumor burden. The standard treatment approach involves a combination of cytoreductive surgery aiming to complete gross resection (CGR) and platinum-based chemotherapy, with or without maintenance therapy \cite{petrillo2015definition}. The sequence of these treatments depends on patient’s characteristics at upfront evaluation. As a matter of fact, if resection of all macroscopic disease can be obtained with an acceptable operative morbidity, the patient undergoes primary cytoreductive surgery (PCS) followed by adjuvant chemotherapy. Otherwise, three to four cycles of neoadjuvant chemotherapy (NACT) are administered and followed by an interval cytoreductive surgery (ICS), if possible \cite{fagotti2011learning}. In this setting, a standardized assessment of disease diffusion is essential to predict the likelihood of CGR. 

The integration of clinical, biochemical, and radiological assessments alone is insufficient to decide on the best strategy between PCS and NACT. In fact, given the low sensitivity of imaging techniques in detecting miliary peritoneal carcinomatosis, diagnostic laparoscopy (DL) still plays a pivotal role by offering direct visualization of peritoneal disease and can also provide a definitive histopathological diagnosis \cite{armstrong2021ovarian,gonzalez2023newly,van2020diagnostic,nougaret2012ovarian}.

Several scoring systems, including the Fagotti score (FS), have been developed to estimate the feasibility of PCS during DL based on the extent of peritoneal carcinomatosis evaluated by the surgeon through direct visualization \cite{avesani2020radiological}. The clinical validity of such scoring systems has been extensively demonstrated in the literature, confirming their predictive value for surgical outcomes and postoperative morbidity \cite{ledermann2024esgo,colombo2019esmo,fagotti2006laparoscopy}.

However, the accuracy and reproducibility of these scoring systems heavily depend on the operator’s expertise, resulting in substantial inter-observer variability \cite{fagotti2008prospective,fagotti2010comparison,moller2025assessing}. Moreover, there are still some grey zones where a clear assessment is difficult to achieve. This reliance on subjective interpretation limits their implementation for intraoperative decision-making during DL, underscoring the need for objective and automated assessment tools for peritoneal carcinomatosis (PC) and surgical feasibility.

In minimally invasive surgery, artificial intelligence (AI) and computer vision (CV) algorithms have been proposed to automatically analyse endoscopic videos and assist intraoperative decision-making, mainly by providing technical guidance such as anatomical structure recognition, phase recognition, and safety checks \cite{mascagni2022computer}. In oncologic settings, recent AI systems have focused on lesion-level tasks, such as distinguishing benign from malignant peritoneal lesions or segmenting intra-abdominal metastases in gastric cancer \cite{schnelldorfer2024development,chen2025artificial}. The present study aims to develop and test an AI model for automated FS assessment on DL videos in patients affected by AOC to provide objective, reproducible decision support during surgery.

\section{Materials and Methods}
\label{sec:methods}

This is a developmental study using retrospectively collected data. The local ethical committee approved the present study (IRB 6854/0022501/24, NCT06017557). Patients signed informed consent to collect their data for research purposes.

\subsection{Data Collection}
\label{subsec:analysis}

Patients undergoing DL for AOC at first diagnosis (FIGO stage IIIB-IVB) at Fondazione Policlinico Universitario Agostino Gemelli IRCCS (Rome, Italy) in 2023 were assessed for eligibility. Exclusion criteria included non-epithelial histology, cases of secondary cytoreductive surgery, patients unfit for surgery, and absence of surgical video recordings. 

DL videos, intraoperative FS assessments, and the relative clinical indications (i.e., indication to PCS or to NACT) were collected retrospectively. Videos were de-identified removing metadata as well as out-of-body images and the whole dataset was pseudo-anonymized \cite{lavanchy2023preserving}.

Data were split according to the date of DL: the first 101 consecutive videos were assigned to a development dataset used for data annotation, AI training, validation and internal testing, whereas the subsequent 50 videos formed an independent test dataset, used exclusively for further internal validation. 

\subsection{Data Annotation}
\label{subsec:delphi_consensus}

Temporal and spatial annotations of videos from the development dataset were performed using MOSaiC, a browser-based platform for collaborative surgical video analysis designed by the IHU-Strasbourg (France) \cite{mazellier2023mosaic}.

First, timestamps were temporally annotated to define the video regions of interest (ROIs) useful for FS assessment (e.g. non-informative phases such as trocar insertion were excluded). Frames were then extracted from the video ROIs at a fixed interval (every five seconds) to obtain a representative dataset and avoid selection biases. In these frames, all anatomical structures contributing to the FS assessment were manually segmented together with PC, including diaphragm, liver, stomach, spleen, lesser omentum, greater omentum, parietal peritoneum, bowel, and PC.

Subsequently, FS assessments performed intraoperatively by operating surgeons and relative clinical indications were retrieved from the operative reports and were associated with each included video, both in the development and independent test dataset. For each case, we recorded the total FS (range 0–12) and the binary status of each FS station (positive = 2 points, negative = 0 points).

For clarity, we distinguished between the individual anatomical structures segmented at the frame level (diaphragm, liver, stomach, spleen, lesser omentum, greater omentum, parietal peritoneum, bowel), and anatomical stations (AS), which correspond to the six FS stations used for video-level classification of PC: diaphragm / liver / stomach-spleen-lesser omentum / greater omentum / parietal peritoneum / bowel. We then derived the Indication to Surgery (ItS) label, defined as a binary clinical variable summarising whether the patient was considered eligible for cytoreductive surgery according to the FS (FS < 8: surgery indicated; FS $\geq$ 8: surgery contraindicated). All these labels defined the clinical ground truth for both the development and independent test datasets. 

Frame-level ROI and segmentation annotations were performed only in the development dataset. In the independent test dataset, no additional frame-level annotations were generated; each video was assigned only the intraoperative FS and the derived ItS label and was reserved exclusively for independent evaluation of video-level FS and ItS predictions. 

Annotators were surgeons experienced in AOC. In addition, annotations were periodically audited by a certified European Society of Gynaecological Oncology (ESGO) surgeon to ensure accuracy and consistency.  

\subsection{AI Training}
\label{subsec:developments_validation_coloworkflow}

\begin{figure*}[tp]
\centering
\includegraphics[width=0.9\textwidth]{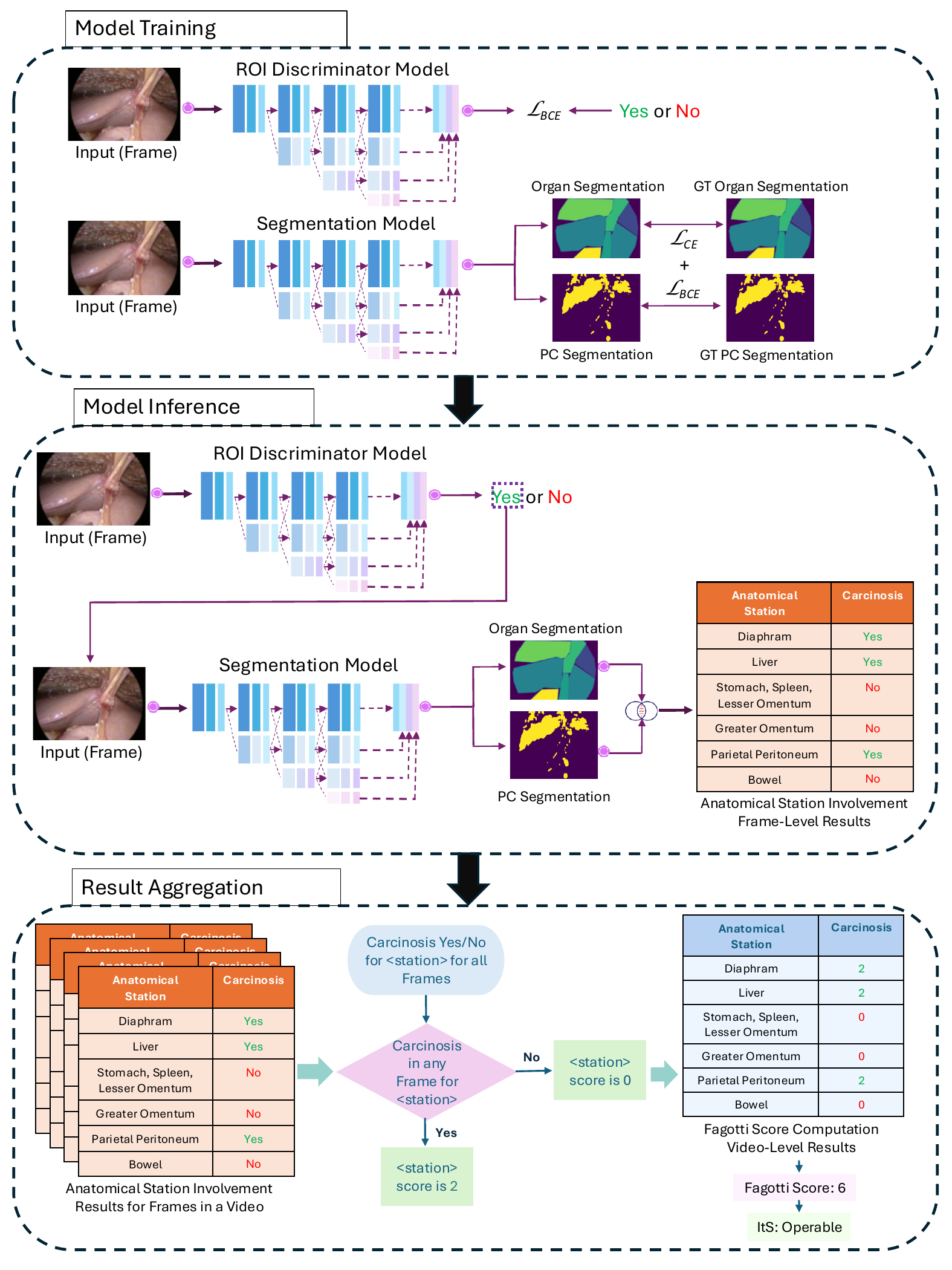}
\caption{Summary of pipeline for Fagotti Score computation for videos. The ROI discriminator selects frames where FS assessment should be performed. Frames extracted from videos are used to train the model for simultaneous organ segmentation and peritoneal carcinosis segmentation. Frame-level anatomical station involvement is derived by detecting peritoneal carcinomatosis nodules within the segmented organ masks belonging to each station. Video- level results are aggregated from frame level results per station. Video-level results are obtained by aggregating frame-level predictions for each anatomical station; each positive anatomical station contributes 2 points to the final FS, which is then used to determine the Indication to Surgery (FS < 8 vs $\geq$ 8).}
\label{fig:figure_1}
\end{figure*}

As shown in \textbf{Fig. \ref{fig:figure_1}.}, the AI pipeline used raw DL videos as input and produced both frame-level and video-level assessments of PC according to the FS. A multi-task deep learning method (DLM) was designed to (1) detect ROIs relevant to the FS assessment and (2) segment anatomical structures and PC. The frame-level predictions from the DLM were then used to classify each organ and each AS as involved or not by PC. The video-level FS was computed by assigning a score to each AS, which was then summed to obtain the final FS for each video. 

The architecture was a Vision Transformer Base model initialized with DINOv2 self-supervised weights \cite{oquab2023dinov2}. The pipeline included two main components: (1) a ROI discriminator, trained using Binary Cross Entropy Loss, to identify video segments relevant for FS assessment; and (2) a segmentation model, trained using a combination of Cross Entropy Loss (for organ segmentation) and Binary Cross Entropy Loss (for PC segmentation). 

At inference, frames selected by the ROI discriminator were processed by the segmentation model to segment anatomical structures and PC. Organ masks with prediction of confidence $\geq$70\% and PC masks of confidence $\geq$90\% were selected while the individual carcinosis nodules were extracted using connected components analysis. Each nodule was then assigned to the anatomical structure with the highest overlapping confidence; in cases of overlap with multiple structures, assignment was based on the organ with the greatest overlap. At the frame level, AS were classified as positive if at least one nodule was detected on the corresponding anatomical structure.   

Frame-level predictions were then aggregated to obtain a single video-level prediction for each AS, which was considered positive if it was ever classified as positive in any frame of the DL and negative otherwise. 

For video-level FS computation, each positive AS was assigned a score of 2 and each negative AS a score of 0. The final FS per video was obtained by summing the scores of the six clinically relevant AS, consistent with the original FS scoring system.  

\section{Experiments and Evaluation Metrics}
\label{sec:results}

The development dataset was divided into four stratified non-overlapping subsets with similar statistical distribution of FS for 4-fold cross validation. Splits were performed at a video level to prevent contamination (i.e., making sure that images from the same case appeared only in one of the training/validation/testing splits). Each fold served as validation and test for one of the 4 cross-validation experiments.  The mean and standard deviation of the test metrics from each of the four cross-validation folds were reported.  For the independent test dataset, each of the four DLMs was evaluated, and the mean and standard deviation of all metrics were reported.  

The proposed AI pipeline was evaluated at multiple levels.  
\begin{enumerate}[label=\arabic*)]
  \item The performance of the ROI frame discriminator was assessed using balanced accuracy, to account for class imbalance between relevant and non-relevant frames.
  \item Frame-level segmentation performance for both anatomical structures and PC was assessed using the Dice similarity coefficient. This analysis was restricted to the development dataset, where manual segmentation masks were available.
  \item Video-level AS classification focused on the six AS included in the FS assessment. For each video and station, prediction of the binary presence or absence of PC (AS involvement) were evaluated using three performance metrics: precision, recall, and F1-score, computed separately in the development and independent test datasets.
  \item Video-level final FS prediction was evaluated by comparing the model’s estimated FS against the intraoperative ground truth assessed by surgeons, using the Root Mean Square Error (RMSE) as the performance metric.
  \item For the ItS metric we computed precision, recall, and F1-score on both the development and independent test datasets.
\end{enumerate}

\section{Results}

\begin{table}[t]
\centering
\caption{Summary of annotated frames in the development dataset. The table reports, for each segmentation label, the number of frames in which the corresponding organ or peritoneal carcinomatosis was annotated, together with the total number of ROI frames.}
\label{tab:frame_no}
\small
\setlength{\tabcolsep}{10pt} 
\begin{tabular}{ll}
\toprule
\textbf{Segmentation label} & \textbf{\begin{tabular}[c]{@{}l@{}}Number of \\ annotated frames\end{tabular}} \\ 
\hline
\textbf{Total ROI frames}   & \textbf{7311}                                                                  \\
Diaphragm                   & 3021                                                                           \\
Liver                       & 3231                                                                           \\
Stomach                     & 1997                                                                           \\
Spleen                      & 273                                                                            \\
Lesser Omentum              & 519                                                                            \\
Greater Omentum             & 4529                                                                           \\
Parietal peritoneum         & 3999                                                                           \\
Bowel                       & 4089                                                                           \\
Peritoneal Carcinomatosis   & 5358                                                                           \\
\bottomrule
\end{tabular}
\end{table}

\begin{table}[t]
\centering
\caption{Frame-level segmentation performance for anatomical structures and peritoneal carcinomatosis in the development dataset, reported as Dice scores (mean ± SD,\%) over the four cross-validation folds.}
\label{tab:seg_results}
\small
\setlength{\tabcolsep}{8pt} 
\begin{tabular}{ll}
\toprule
\textbf{Organ} & \textbf{Dice Score (\%)} \\ 
\hline
Diaphragm & 72 ± 4 \\
Liver & 87 ± 2\\
Stomach & 74 ± 2 \\
Spleen & 55 ± 16 \\
Lesser Omentum & 39 ± 5 \\
Greater Omentum & 80 ± 1 \\
Parietal peritoneum & 76 ± 2 \\
Bowel & 74 ± 2 \\
\textbf{Average for anatomical structures} & \textbf{70 ± 3}\\
\textbf{Peritoneal Carcinomatosis}   & \textbf{56 ± 3} \\
\bottomrule
\end{tabular}
\end{table}

\begin{table}[t]
\centering
\caption{Video-level anatomical station involvement and Indication to Surgery (ItS) classification performance in the development dataset and independent test dataset. 
For each anatomical station and ItS category, precision, recall, and F1-score are reported as mean ± standard deviation (\%), computed over the four cross-validation models.}
\label{tab:station_results}
\small
\setlength{\tabcolsep}{3pt} 
\begin{tabular}{llll}
\toprule
\multicolumn{4}{l}{\textbf{Development dataset (101 videos)}} \\
\hline
\textbf{AS Involvement} & \textbf{Precision} & \textbf{Recall} & \textbf{F1-score} \\ 
\hline
Diaphragm & 91 ± 6 & 93 ± 5 & 92 ± 5 \\
Liver & 48 ± 10 & 51 ± 21 & 49 ± 14 \\
Stomach, Spleen, Lesser Omentum & 76 ± 21 & 38 ± 11 & 51 ± 13 \\
Greater Omentum & 89 ± 8 & 87 ± 3 & 88 ± 5 \\
Parietal peritoneum & 77 ± 5 & 91 ± 9 & 83 ± 7 \\
Bowel & 72 ± 11 & 89 ± 4 & 79 ± 5 \\
\textbf{AS Involvement Average} & \textbf{76 ± 1} & \textbf{75 ± 5} & \textbf{74 ± 3}\\
\midrule
ItS < 8 & 80 ± 10 & 78 ± 11 & 79 ± 10 \\
ItS $\geq$ 8 & 82 ± 8 & 83 ± 8 & 82 ± 8 \\
\textbf{ItS Average} & \textbf{82 ± 8} & \textbf{83 ± 8} & \textbf{82 ± 8} \\
\midrule
\multicolumn{4}{l}{\textbf{Independent test dataset (50 videos)}} \\
\hline
\textbf{AS Involvement} & \textbf{Precision} & \textbf{Recall} & \textbf{F1-score} \\ 
\hline
Diaphragm & 96 ± 3 & 95 ± 4 & 95 ± 3 \\
Liver & 48 ± 10 & 41 ± 6 & 44 ± 6 \\
Stomach, Spleen, Lesser Omentum & 67 ± 24 & 43 ± 22 & 49 ± 19 \\
Greater Omentum & 92 ± 3 & 87 ± 6 & 89 ± 3 \\
Parietal peritoneum & 91 ± 1 & 89 ± 7 & 90 ± 3 \\
Bowel & 70 ± 3 & 80 ± 7 & 75 ± 3 \\
\textbf{AS Involvement Average} & \textbf{77 ± 3} & \textbf{72 ± 7} & \textbf{73 ± 4}\\
\midrule
ItS < 8 & 84 ± 1 & 80 ± 5 & 82 ± 3 \\
ItS $\geq$ 8 & 76 ± 5 & 81 ± 3 & 78 ± 2 \\
\textbf{ItS Average} & \textbf{80 ± 2} & \textbf{80 ± 2} & \textbf{80 ± 2} \\
\bottomrule
\end{tabular}
\end{table}

\begin{figure*}[tp]
\centering
\includegraphics[width=0.8\textwidth]{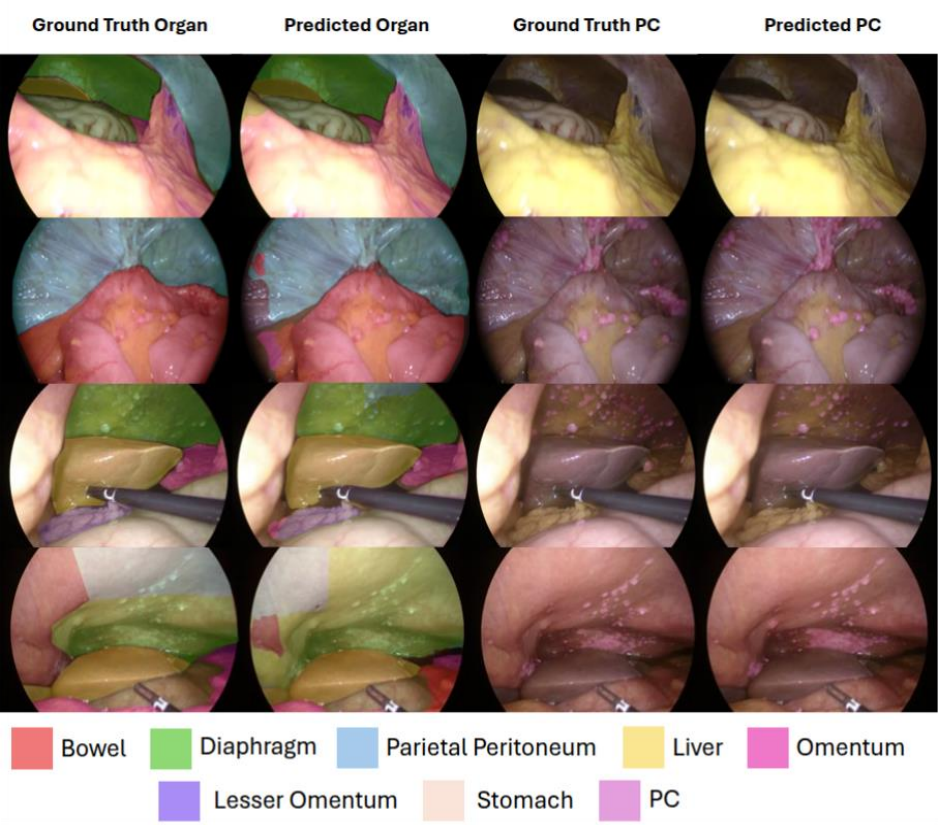}
\caption{Qualitative results of the model outputs compared with the annotated ground truth.}
\label{fig:figure_2}
\end{figure*}

\subsection{Development Dataset}
From the ROIs of the 101 videos in the development dataset, 7,311 frames were extracted (\textbf{Table \ref{tab:frame_no}.}).  

The ROI frame discriminator achieved a balanced accuracy of 86±1.41\%.  

Frame-level segmentation performance showed a mean Dice score of 70±3\% for anatomical structures and 56±3\% for PC. Dice scores varied by structures: the liver (87±2\%) and greater omentum (80±1\%) showed the highest accuracy, whereas the lesser omentum (39±5\%) and spleen exhibited lower values (\textbf{Table \ref{tab:seg_results}}). \textbf{Fig. \ref{fig:figure_2}.} illustrates examples of the model’s performance, providing both static visualizations of anatomical and PC segmentation during DL.

At video-level, AS involvement classification performance was highest for the diaphragm (F1-score: 92±5\%), greater omentum (88±5\%), and parietal peritoneum (83±7\%). Lower F1-scores were observed in the station of the stomach, spleen, and lesser omentum (51±13\%), as well as in the liver (49±14\%). The average F1-score across all stations was 74±3\% (Table 3). 

Video-level final FS prediction showed a root mean square error (RMSE) of 2.78±0.35 points compared to the ground-truth assessments by surgeons. Considering that the FS is scored in 2-point steps, this corresponds to an average error of one FS level (normalized RMSE of 1.39± 0.18). For the ItS classification, the model achieved an overall F1-score of 80±8\%. F1-score for operable cases (FS<8) was 79±10\% and for inoperable ones (FS $\geq$ 8) 82±8\% (\textbf{Table \ref{tab:station_results}}).

\subsection{Independent Test Dataset}

In the independent test dataset (50 videos), AS classification performance was highest for the diaphragm (F1-score: 95±3\%), followed by the parietal peritoneum (90±3\%) and greater omentum (89±3\%). Lower F1-scores were found in the stomach/spleen/lesser omentum (49±19\%) and liver (44±6\%) stations. The overall average F1-score across stations was 73±4\% (\textbf{Table \ref{tab:station_results}}).

The RMSE for the predicted video-level FS was 2.29±0.15, compared with 2.78±0.35 in the development dataset. The normalized RMSE in the independent test dataset was 1.15±0.08.

For the ItS prediction, the model achieved an overall F1-score of 80±2\% (FS<8: F1-score 82±3\%; FS$\geq$8: F1-score 78±2\%) (\textbf{Table \ref{tab:station_results}}).

\section{Discussion}
\label{sec:discussion}

To the best of our knowledge, this is the first study to demonstrate the application of AI for the automatic assessment of PC during DL for AOC. This system aims to support intraoperative decision-making and help clinicians to achieve a more reliable and consistent management of women affected by this disease.  

Recent studies by Schnelldorfer et al. and Chen et al. have paved the way for the use of AI in the assessment of PC, particularly in gastrointestinal cancer, demonstrating the feasibility and scientific relevance of this approach. Schnelldorfer et al. developed a model which differentiated malignant from benign lesions in staging laparoscopy images, while Chen et al. demonstrated real-time segmentation of intra-abdominal metastases in gastric cancer surgery \cite{schnelldorfer2024development, chen2025artificial}. Notably, these systems focus on lesion recognition and do not quantify disease burden using validated scoring systems such as the FS or Peritoneal Cancer Index (PCI) \cite{jacquet1996clinical}.

Our model is based on a CV algorithm which automatically assesses FS during DL, aiming to reduce subjectivity and enhance accuracy in one of the most operator-dependent aspects of AOC management \cite{moller2025assessing}.

The model was evaluated through four-fold cross-validation on the development dataset and validated on an independent test dataset, performing robustly across both datasets. This independent test dataset, although limited in size, served as a dedicated internal validation cohort, allowing us to assess model performance on entirely unseen data and verify its generalizability beyond the training pipeline. 

Frame level segmentation achieved a mean Dice score of 70\% for organ segmentation and 56\% for PC segmentation, with minimal score deviation (RMSE around 2 points) from the clinical referral value (given the FS increments by two per each scoring step).  

Station-level results demonstrate the usefulness and pitfalls of AI in accurately identifying PC in key AS. Performance was high in well-exposed regions such as the diaphragm, parietal peritoneum, and omentum. Conversely, performance was lower in more anatomically complex or less visible areas such as spleen and lesser omentum. These results reflect both real-world clinical aspects of intraoperative visualization and annotation variability, underscoring the need for larger and more heterogeneous datasets to enhance generalizability. 

Overall, ItS performance averaged an F1-score of 80\% both in the development datasets and in the independent test datasets, thereby demonstrating consistent performance in the model’s validation cohort. 

Despite its results, this study presents several limitations.

\begin{enumerate}[label=\alph*)]
  \item Mesenteric retraction and miliary small bowel carcinomatosis, both criteria for inoperability, were difficult to annotate in our dataset, given their relatively low incidence in real-life clinical practice, limiting the model’s exposure and subsequent prediction of these conditions.
  \item The lower accuracy in detecting PC in some difficult-to-reach anatomical structures (e.g., stomach, spleen, lesser omentum) may be related to the retrospective nature of this study where these regions were not extensively registered. The integration of preoperative imaging data has the potential to mitigate this risk together with larger and heterogeneous datasets allowing the training of DLM on underrepresented anatomical structures.
  \item External validation across multiple institutions will be critical to improve the robustness of the model.
\end{enumerate}

Despite these limitations, this study represents the proof of concept that intraoperative AI can be effectively employed during DL in AOC with the need for continuous iteration, prospective validation and extension of the model. 

One of the next steps will be expanding the applicability of this model to other intraoperative scoring systems such as PCI, which also relies on the topographic assessment of PC. Indeed, the application of our model to PCI assessments by adapting the same segmentation and identification pipeline would be a significant step towards its generalization and scalability. Moreover, the ability to decouple the CV algorithm from a fixed score and predefined anatomical regions, focusing instead solely on the final outcome, could unveil unexpected capabilities of AI.  

Future developments of this approach may also include the correlation of intraoperative macroscopic visual features associated with tumor biology such as histologic subtypes, molecular alterations (e.g., BRCA status, HRD/HRP), or expected response to NACT, ultimately enhancing individualized treatment decisions.  

Finally, clinical translation of this AI system will be pivotal. Mascagni et al. showed that AI models can run in the OR, analyzing the laparoscopic video feed in real-time to overlay segmentation masks and provide feedback during surgery \cite{mascagni2024early}. A similar approach could be envisioned to integrate the present DLM during DL for AOC. In early clinical studies, surgeons would be blinded, with the AI deployed in a “shadow mode” to verify its reliability without directly influencing surgical decision making.

\section{Conclusions}

This study represents a significant leap forward in leveraging AI for the management of AOC. The model’s success in predicting FS, combined with high segmentation accuracy in key anatomical structures, demonstrates its potential to democratize intraoperative clinical decision-making and finally optimize cytoreductive outcomes. Despite challenges in achieving consistent segmentation across all anatomical regions, the results indicate a clear path forward for refining the algorithm and expanding its clinical applicability. Future studies will aim to enhance the model's precision in complex anatomical regions and integrate histological, molecular and genetic data to further tailor treatment strategies for AOC patients. 

\section{Disclosures}

Pietro Mascagni and Nicolas Padoy are co-founders and shareholders of Scialytics. The other co-authors do not have any conflict of interests to disclose.

\section{Acknowledgement}

The authors thank the CAMMA/IHU MOSaiC team for their technical support with the video annotation platform MOSaiC. Their gratitude is also extended to the patients and surgical teams at Fondazione Policlinico Universitario A. Gemelli IRCCS for their contributions to data collection. This work has received funding from the European Union (ERC, CompSURG, 101088553).  The views and opinions expressed are, however, those of the authors only and do not necessarily reflect the views of the European Union or the European Research Council. Neither the European Union nor the granting authority can be held accountable for them. This work has also been supported by French state funds managed within the Plan Investissements d’Avenir by the Agence Nationale de la Recherche (ANR) under reference ANR- 10-IAHU-02 (IHU Strasbourg). This work was granted access to the servers/HPC resources managed by CAMMA, IHU Strasbourg, Unistra Mesocentre, and GENCI-IDRIS [Grant 2021- AD011011638R3, 2021-AD011011638R4].

\bibliographystyle{elsarticle-num}
\bibliography{arxiv}

\end{document}